# Type variables in patterns

Richard A. Eisenberg
Bryn Mawr College
Bryn Mawr, PA, USA
rae@cs.brynmawr.edu

Joachim Breitner
University of Pennsylvania
Philadelphia, PA, USA
joachim@cis.upenn.edu

Simon Peyton Jones
Microsoft Research
Cambridge, UK
simonpj@microsoft.com

## Abstract

For many years, GHC has implemented an extension to Haskell that allows type variables to be bound in type signatures and patterns, and to scope over terms. This extension was never properly specified. We rectify that oversight here. With the formal specification in hand, the otherwise-labyrinthine path toward a design for binding type variables in patterns becomes blindingly clear. We thus extend *ScopedTypeVariables* to bind type variables explicitly, obviating the *Proxy* workaround to the dustbin of history.

## 1 Introduction

Haskell allows the programmer to write a *type signature* for a definition or expression, both as machine-checked documentation, and to resolve ambiguity (Section 2.1). For example,

$$prefix :: a \rightarrow [[a]] \rightarrow [[a]]$$
$$prefix\ x\ yss = map\ xcons\ yss$$
$$\quad \textbf{where}\ xcons\ ys = x : ys$$

Sadly, it is not always possible to write such a type signature. For example, to give a signature for *xcons* we might try:

$$prefix :: a \rightarrow [[a]] \rightarrow [[a]]$$
$$prefix\ x\ yss = map\ xcons\ yss$$
$$\quad \textbf{where}\ xcons :: [a] \rightarrow [a]$$
$$\quad\quad\quad xcons\ ys = x : ys$$

But Haskell98's scoping rules specify that the *a* in the signature for *xcons* is *locally* quantified, thus: $xcons :: \forall a. [a] \rightarrow [a]$. That is not what we want! We want the *a* in the signature for *xcons* to mean "the universally quantified type variable for *prefix*", and *Haskell98 provides no way to do that.*

The inability to supply a type signature for *xcons* might seem merely inconvenient, but it is just the tip of an iceberg. Haskell uses type inference to infer types, and that is a wonderful thing. However, insisting on *complete* type inference—that is, the ability to infer types for any well-typed program with no help from the programmer—places serious limits on the expressiveness of the type system. GHC's version of Haskell has gone far beyond these limits, and fundamentally relies on programmer-supplied annotations to guide type inference. As some examples, see the work of Peyton Jones et al. [2007], Vytiniotis et al. [2011], or Eisenberg et al. [2016].

So the challenge we address is this: *it should be possible for the programmer to write an explicit type signature for any sub-term of the program.* To do so, some type signatures must refer to a type that is already in the static environment, so we need a way to *name* such types. The obvious way to address this challenge is by providing language support for *lexically scoped type variables.* GHC has long supported scoped type variables: the *ScopedTypeVariables* extension is very popular, and 29% of Haskell packages on Hackage use it. But it has never been formally specified! Moreover, as we shall see, it is in any case inadequate to the task. In this paper we fix both problems, making the following contributions:

- In the days of Haskell98, scoped type variables were seldom crucial. Through a series of examples we show that, as Haskell's type system has grown more sophisticated, the need for scoped type variables has become acute (Section 2), while at the same time GHC's existing support for them has become more visibly inadequate (Section 3).
- To fix these inadequacies, we describe *visible type application in patterns*, a natural extension to GHC's existing visible type applications from terms to patterns (Section 4).
- We give the first formal specification of scoped type variables for Haskell, formalizing the folklore, and providing a firm foundation for both design and implementation (Section 5).
- As part of this specification, we offer a new and simpler typing judgment for GADT pattern matching (Section 5.3), which treats uniformly the universal and existential variables of a data constructor.

## 2 Motivation and background

### 2.1 The need for type annotations

One of the magical properties of ML-family languages, including Haskell, is that type inference allows us to write many programs with no type annotations whatsoever. In practice, however, Haskell programs contain many user-written type signatures, for two main reasons.

First, the type of a function can be extremely helpful as documentation, with the advantage that it is *machine-checked* documentation. Almost all programmers regard it as good practice to provide a signature for every top-level function. Indeed, GHC has a warning flag, `-Wmissing-signatures`, which enforces this convention.

```
show   :: Show a ⇒ a → String
read   :: Read a ⇒ String → a
(+)    :: [a] → [a] → [a]
concat :: [[a]] → [a]
```

**Figure 1.** Types of standard functions

Second, as Haskell's type system becomes increasingly expressive, complete type inference becomes intractable, and the type system necessarily relies on programmer-supplied type annotations. Here are some examples:

- *Type-class ambiguity* is present even in Haskell 98. Consider[1]:

  ```
  normalize :: String → String
  normalize s = show (read s)
  ```

  This function parses a string to a value of some type, and then turns that value back into a string. But nothing in the code specifies that type, so the programmer must disambiguate. One way to do so is to provide a type signature that specifies the result type of the *read s* call, thus:

  ```
  normalize s = show (read s :: Int)
  ```

- *Polymorphic recursion.* In ML, recursive calls to a function must be at monomorphic type, but Haskell has always supported polymorphic recursion, provided the function has a type signature. For example:

  ```
  data T a = Leaf a | Node (T [a]) (T [a])
  leaves :: T a → [a]
  leaves (Leaf x)    = [x]
  leaves (Node t1 t2) = concat (leaves t1 ++ leaves t2)
  ```

- *Higher-rank types* [Peyton Jones et al. 2007]. Consider

  ```
  f :: (∀a. [a] → [a]) → ([Char], [Bool])
  f g = (g "Hello", g [True, False])
  ```

  Here the type of *g* is polymorphic, so it can be applied to lists of different type. The type signature is essential to specify the type of the argument *g*; without it, *f* will be rejected.

- *Generalized algebraic data types* [Schrijvers et al. 2009]. The popular *GADTs* extension to GHC allows pattern matching to refine the type information available in the right hand side of an equation. Here is an example:

  ```
  data G a where
    MkInt :: G Int
    MkFun :: G (Int → Int)
  ```

  When we learn that a value *g :: G a* is actually the constructor *MkInt*, then we simultaneously learn that

  ---
  [1] Figure 1 gives the types of standard functions, such as *read* and *show*.

*a* really is *Int*. GHC can use this fact during type checking the right-hand side of a function, like this:

```
matchG :: G a → a
matchG MkInt = 5
matchG MkFun = (+10)
```

Again, however, *matchG* will only type-check if it is given a signature; see Schrijvers et al. [2009] for details.

- *Ambiguous types.* Consider

  ```
  type family F a
  ambig :: Typeable a ⇒ F a → Int
  test :: Char → Int
  test x = ambig x
  ```

  In *test* GHC must decide at what type to call *ambig*; that is, what type should instantiate the *a* in *ambig*'s type. Any choice *a = τ* must ensure that *F τ ∼ Char* but, because *F* might not be injective, that does not tell us what *a* should be. A type signature is not enough to resolve this case; we need a different form of type annotation, namely *visible type application* (Section 2.3).

There is a general pattern here: as the type system becomes more expressive, the type inference needs more guidance. Moreover, that guidance is extremely informative to the programmer, as well as to the compiler.

## 2.2 Support for scoped type variables

Given the increasing importance of type annotations, a good principle is this: it should be possible to specify, via a type signature, the type of any sub-expression or any let-binding. Alas, as shown in the introduction, Haskell98 supports only *closed* type signatures, so there are useful type signatures that we simply cannot write.

The key deficiency in Haskell98 is that it provides no way to bring type variables into scope. GHC has recognized this need for many years, and the *ScopedTypeVariables* extension offers two ways to bring a type variable into scope:

- *Binding in a declaration signature.* Since 2004 GHC allows you to write

  ```
  prefix :: ∀a. a → [[a]] → [[a]]
  prefix x yss = map xcons yss
      where xcons :: [a] → [a]
            xcons ys = x : ys
  ```

  The explicit "∀" brings *a* into scope in the rest of the type signature (of course), but it *also* brings *a* into scope in the body of the named function, *prefix* in this case. The rule is a bit strange, because the definition of *prefix* is not syntactically "under" the ∀, and indeed the signature can be written far away from the actual binding. But in practice the rule works well, and we take it as-is for the purposes of this paper.

- *Pattern signatures*: binding a type variable in a pattern. For even longer, since 1998, GHC has allowed you to write this:

```
prefix (x :: b) yss = map xcons yss
  where xcons :: [b] → [b]
        xcons ys = x : ys
```

Here, we bind the type variable *b* in the pattern (*x :: b*), and this binding scopes over the body of the binding. We describe pattern signatures in much more detail in Section 3.

### 2.3  Visible type application

The *TypeApplications* extension provides a relatively new form of type annotation: explicit type applications, first described by Eisenberg et al. [2016]. The idea is that an argument of the form @*ty* specifies a *type* argument. This can often be used more elegantly than the body of a type signature. For example, a hypothetical unit-test for the function *isJust* :: *Maybe a → Bool*,

```
testIsJust1 = isJust (Just (2018 :: Int))     == True
testIsJust2 = isJust (Nothing :: Maybe Int) == False
```

can equivalently be written more elegantly using explicit type annotations

```
testIsJust1 = isJust (Just @Int 2018) == True
testIsJust2 = isJust (Nothing @Int)   == False
```

Visible type application solves the awkward case of *ambig* in Section 2.1: we can disambiguate the call with a type argument. For example:

```
type family F a
type instance F Bool = Char

ambig :: Typeable a ⇒ F a → Int

test :: Char → Int
test x = ambig @Bool x
```

Here we specify that *ambig* should be called at *Bool*, and that is enough to type-check the program.

It is natural to wonder whether we can extend visible type application to patterns, just as we extended type signatures to patterns. Doing so is the main language extension suggested in this paper: Section 4.

## 3  Pattern signatures and their shortcomings

We see above that *ScopedTypeVariables* enables the user to bind type variables in patterns, by providing a *pattern signature*, that is, a type signature in a pattern. We explore pattern signatures and their shortcomings, in this section.

### 3.1  The binding structure of a pattern signature

A pattern signature may *bind* a type variable, but it may also *mention* a type variable that is already in scope. For example, we may write

```
prefix (x :: a) yss = map xcons yss
  where xcons (ys :: [a]) = x : ys
```

Here, the pattern signature (*x :: a*) binds *a* (as well as *x*), but the pattern signature (*ys :: [a]*) simply mentions *a* (which is already in scope), as well as binding *ys*. The rule is this: a use of a type variable *p* in a pattern signature is an *occurrence* of *p* if *p* is already in scope; but *binds p* if *p* is not already in scope.

It is entirely possible to have many different type variables in scope, all of which are aliases for the same type. For example:

```
prefix :: ∀a. a → [[a]] → [[a]]
prefix (x :: b) (yss :: [[c]]) = map xcons yss
  where xcons (ys :: [d]) = x : ys
```

Here *a*, *b*, *c*, and *d* are all in scope in the body of *xcons*, and are all aliases for the same type.

The current implementation of *ScopedTypeVariables* allows such lexically-scoped type variables to stand only for other *type variables*, and not for arbitrary *types*, a point we return to in Section 3.5.

### 3.2  Pattern signatures are useful

Pattern signatures have merit even if there are no type variables around. Consider this Haskell program:

```
main = do x ← readLn
          if null x then putStrLn "Empty"
                    else putStrLn "Not empty"
```

The types of the program are ambiguous: Clearly, *x* is some type that has a *Read* instance, and because it is passed to *null*, is a list[2], but the compiler needs to know the precise type, and rejects the program.

To fix this in Haskell98, the programmer has two options:

- They can wrap the call to *readLn* in a type annotation:

```
main = do x ← (readLn :: IO [Int])
          if null x then putStrLn "Empty!"
                    else print x
```

but this is infelicitous because there is no question that *readLn* is in the *IO*, and with larger types this can get very verbose.

- They can wrap an *occurrence* of *x* in a type annotation:

```
main = do x ← readLn
          if null (x :: [Int]) then putStrLn "Empty"
                               else print x
```

---

[2] Or, with a recent version of the standard library, it is something with a `Foldable` instance.

but again this is unsatisfying, because it feels too late. Both variants are essentially work-arounds for the natural way of specifying the type of *x*, namely at its binding site:

```
main = do (x :: [ Int ]) ← readLn
          if null x then putStrLn "Empty"
                    else print x
```

which is precisely what the *PatternSignatures* language extension provides[3]—the ability to write a type annotation in a pattern.

Users, especially beginners, who have to track down a confusing type error in their code, can now exhaustively type annotate not just their terms, but also their patterns, until they have cornered the bug.

### 3.3 Pattern signatures are essential

Pattern signatures become more crucial when we consider existential types. The *ExistentialQuantification* extension allows users to bind *existential* variables in their data constructors. These are type variables whose values are "stored" by a constructor (but not really, because types are erased) and made available during pattern matching. Here are two examples:

```
data Ticker where
   MkTicker :: ∀a. a → (a → a) → (a → Int) → Ticker

data Showable where
   MkShowable :: ∀a. Show a ⇒ a → Showable
```

A *Ticker* contains an object of some type (but we do not know *what* type), along with an update function of type *a* → *a* and a way to convert an *a* into an *Int*. The *Showable* type packs a value of an arbitrary type that has a *Show* instance along with its *Show* dictionary. Here are some functions that operate on these types:

```
-- Updates a ticker, returning whether or not the ticker
-- has reached a limit
tick :: Ticker → Int → (Ticker, Bool)
tick (MkTicker val upd toInt) limit
   = (newTicker, toInt newVal ⩾ limit)
   where newVal    = upd val
         newTicker = MkTicker newVal upd toInt

showAll :: [ Showable ] → String
showAll [ ]                   = ""
showAll (MkShowable x : ss) = show x ⧺ showAll ss
```

We see that the *tick* function can unpack the existential in the *Ticker* value and operate on the value of type *a* without ever knowing what *a* is. Similarly, the *showAll* function works with data of type *a* knowing only that *a* has a *Show* instance.

---

[3]Modern GHC actually folds *PatternSignatures* into *ScopedTypeVariables*, giving both extensions the same meaning. However, it is expositionally cleaner to separate the two, as we do throughout this paper.

(The *Show a* constraint is brought into scope by the pattern match.)

However, existentials can never escape, forbidding the following function:

```
jailbreak (MkTicker val _ _) = val
```

What should the type of *jailbreak* be? There is no answer to this question (*jailbreak* :: *Ticker* → *a* is clearly too polymorphic), and so GHC rejects this definition, correctly stating that type variable 'a' would escape its scope.

We naturally wish to *name* the existential type variable sometimes. For example, suppose we wanted to give *newVal* (in the **where** clause of *tick*) a type signature. Saying *newVal*:: *a* would hardly do, because there is not yet a connection between the name *a* and the type unpacked from *MkTicker*. We have to do this:

```
tick (MkTicker (val :: b) upd toInt) limit = ...
   where newVal :: b
         newVal = upd val
```

The *val* :: *b* in the pattern binds the type variable *b*, so we can refer back to it later.

### 3.4 Pattern signatures are clumsy

Pattern signatures can be clumsy to use, when the type variable is buried deep inside an ornate type. Here is a contrived example:

```
data Elab where
   MkElab :: Show a ⇒ [ Maybe (Tree (a, Int)) ] → Elab

getE :: Elab → Int
getE (MkElab (xs :: [ Maybe (Tree (a, Int)) ])) = ...a ...
```

To bring *a* into scope in *f*'s right-hand side, we have to repeat the *MkElab*'s elaborate argument type.

More seriously, it may be impossible, rather than merely clumsy, to bind the variable we need. Consider the following GADT:

```
data GM a where
   MkMaybe :: GM (Maybe b)

matchGM :: a → GM a → Bool
matchGM x MkMaybe = isJust x
```

This definition works just fine: GHC learns that *x*'s type is *Maybe b* (for some existential *b*) and the call to *isJust* is well typed. But what if we want to bind *b* in this definition? Annoyingly, *MkMaybe* has no argument to which we can apply a pattern signature. Nor does it work to wrap a pattern signature around the *outside* of the match, thus:

```
matchGM :: a → GM a → Bool
matchGM x (MkMaybe :: GM (Maybe b)) = isJust @b x
```

This definition is rejected. The problem is that the type annotation on the *MkMaybe* pattern is checked *before* the pattern itself is matched against[4]. Before matching the *MkMaybe*, we do not yet know that *a* is really *Maybe b*. Nor can we put the type annotation on *x*, as that, too, occurs before the *MkMaybe* pattern has been matched. A possible solution is this monstrosity:

```
matchGM :: a → GM a → Bool
matchGM x gm@MkMaybe = case gm of
    (_ :: GM (Maybe b)) → isJust @b x
```

This is grotesque. We must do better, and we will in Section 4.

### 3.5 Pattern signatures resist refactoring

In Section 3.1, we explained that a scoped type variable may refer only to another type variable. This means that the definition

```
prefix :: a → [[a]] → [[a]]
prefix (x :: b) yss = map xcons yss
    where xcons ys = x : ys
```

is accepted, because *b* stands for the type variable in the type of *prefix*. But suppose, for example, we specialize the type signature of *prefix* without changing its definition:

```
prefix :: Int → [[Int]] → [[Int]]
prefix (x :: b) yss = map xcons yss
    where xcons ys = x : ys
```

Now this definition is rejected with the error message "Couldn't match expected type *b* with actual type *Int*", because *b* would have to stand for *Int*.

Since the design of *ScopedTypeVariables*, GHC has evolved, and with the advent of type equalities, the restriction itself becomes confusing. Should GHC accept the following definition?

```
prefix :: a ∼ Int ⇒ Int → [[a]] → [[a]]
prefix (x :: b) yss = map xcons yss
    where xcons ys = x : ys
```

Is *b* an alias for *a* (legal) or for *Int* (illegal)? Since *a* and *Int* are equal, the question does not really make sense. We therefore propose to simply drop this restriction (Section 4.3).

### 3.6 Pattern signatures are inadequate

We end our growing list of infelicities with a case in which there is no way whatsoever to bind the type variable, short of changing the data type definition.

Type families [Chakravarty et al. 2005; Eisenberg et al. 2014] allow users to write type-level functions and encode type-level computation. For example, we might write this:

```
type family F a where
    F Int   = Bool
    F Char  = Double
    F Float = Double
```

Naturally, we can use a type family to define the type of an argument to an existential data constructor:

```
data TF where
    MkTF :: ∀a. Typeable a ⇒ F a → TF
```

The *MkTF* constructor stores a value of type (*F a*); it also stores a dictionary for a *Typeable a* constraint [Peyton Jones et al. 2016]—that is, we can use a runtime type test to discover the choice for *a*. We would thus like to write the following function:

```
toDouble :: TF → Double
toDouble (MkTF x)   -- We expect x :: F a
    | Just HRefl ← isType @Int   = if x then 1.0 else − 1.0
    | Just HRefl ← isType @Char = x
    | Just HRefl ← isType @Float = x
    | otherwise                 = 0.0
    where
        isType :: ∀ty. Typeable ty ⇒ Maybe (a :≈: ty)
        isType = eqTypeRep (typeRep @a) (typeRep @ty)
```

The specifics of this function are not important here (see [Peyton Jones et al. 2016]). For our present purposes, the crucial point is this: the existentially-bound type *a* is mentioned in both the definition of *isType* and its type signature—*but there is no way to bring a into scope*. We might try using a pattern signature at the binding of *x*, thus:

```
toDouble (MkTF (x :: F a)) = ...
```

but that does not quite work. The problem is that *F* is not injective. The *a* in that pattern type annotation need not be the same one packed into the existential type variable by *MkTF*, and GHC rightly considers such an *a* to be ambiguous[5].

The only workaround available today is to change *MkTF* to take a proxy argument:

```
data Proxy a = Proxy   -- in GHC's Data.Proxy
data TF where
    MkTF :: ∀a. Typeable a ⇒ Proxy a → F a → TF

toDouble (MkTF (_ :: Proxy a) x) = ...
```

The *Proxy* type stores no runtime information (at runtime, it is isomorphic to ()), but the type *Proxy a* carries the all-important type variable *a*. All datatypes are injective, so we can use this proxy argument to bind the type variable *a* in a way that we could not do previously.

As with many other examples, this is once again unsatisfying: it is a shame that we have to modify the data constructor declaration just to deal with type variable binding.

---

[4]This ordering arises because any type variable in a pattern signature is bound within the pattern.

[5]If *F* were an injective type family, we could label it as such to fix the problem [Stolarek et al. 2015]. But here we assume that it is not.

### 3.7 Conclusion

In this section we have seen that pattern signatures allow us to bring into scope the existentially-bound type variables of a data constructor, but that doing so can be clumsy, and occasionally impossible. We need something better.

## 4 Visible type application in patterns

Consider again *Elab* from Section 3.4:

**data** *Elab* **where**
    *MkElab* :: *Show a* ⇒ [ *Maybe* (*Tree* (*a*, *Int*))] → *Elab*

and suppose we want to build a value of type *Elab* containing an empty list. We cannot write just *MkElab* [ ] because that is ambiguous: we must fix the type at which *MkElab* is called so that the compiler can pick the right *Show* dictionary. We can use a type signature, but it is clumsy, just as the pattern signature was clumsy in Section 3.4:

*MkElab* ([ ] :: [ *Maybe* (*Tree* (*Bool*, *Int*))])

It is much nicer to use visible type application and write *MkElab* @*Bool* [ ]. So it is natural to ask whether we could do the same in patterns, like this:

*getE* :: *Elab* → *Int*
*getE* (*MkElab* @*a* *xs*) = …*a* …

Here, we bind *a* directly, as a type-argument pattern all by itself, rather than indirectly via a pattern signature.

We call this *visible type application in patterns*, a dual of visible type application in the same way that a pattern signatures are a dual of type signatures. This section describes visible type application in patterns informally, while the next formalizes it.

This feature was first requested more than two years ago.[6] Furthermore, binding type variables like this is useful for more than just disambiguation, as we will shortly see.

### 4.1 Examples

Visible type application in patterns immediately fixes the other problems of pattern signatures identified above. For example, in the GADT example of Section 3.4 we can write

*matchGM* :: *a* → *GM a* → *Bool*
*matchGM* *x* (*MkMaybe* @*b*) = *isJust* @*b* *x*

and for the type-family example of Section 3.6 we write

*toDouble* (*MkTF* @*a* *x*) = …

### 4.2 Universal and existential variables

Visible type applications in patterns can be used for *all* the type arguments of a data constructor, whether existential or universal. As an example of the latter we may write

*main* = **do** (*Just* @*Int* *x*) ← *readMaybe* `fmap` *getLine*
        *putStrLn* "Input was " + *show x*

| Term variables | ∋ x, y, z, f, g, h |
|---|---|
| Internal type vars | ∋ a, b |
| User type vars | ∋ c |
| Data constructors | ∋ K |
| Atoms | $v ::= K \mid x$ |
| Expressions | $e ::= v \mid \lambda x.e \mid e_1\ e_2 \mid$ **case** $e$ **of** $\{\overline{p \rightarrow e}\}$ |
| Patterns | $p ::= x \mid K\ \overline{p}$ |
| Polytypes | $\sigma ::= \forall \overline{a}.\ \tau$ |
| Monotypes | $\tau, v ::= tv \mid T\ \overline{\tau} \mid \ldots$ |
| Type variables | $tv ::= a \mid c$ |
| Type env | $\Gamma ::= \epsilon \mid \Gamma, v : \sigma$ |
| Substitutions | $\theta ::= [\overline{tv \mapsto \tau}]$ |
| Types of data cons | $\Gamma_0 = \overline{K : \forall \overline{a}.\ \overline{\sigma} \rightarrow T\ \overline{a}}$ |

**Figure 2.** The initial grammar

as an alternative to

*main* = **do** (*Just* (*x* :: *Int*)) ← *readMaybe* `fmap` *getLine*
        *putStrLn* $ "Input was " + *show x*

Visible type application in patterns considers the type of data constructor, *exactly as written by the user*. For example

**data** *G a b* **where**
    *G1* :: ∀*b*. *Char* → *G Int b*
    *G2* :: ∀*p q b*. *p* → *q* → *b* → *G* (*p*, *q*) *b*
    *G3* :: ∀*p q a b*. (*a* ∼ (*p*, *q*)) ⇒ *p* → *q* → *b* → *G a b*

*f* :: *G a Bool* → *Int*
*f* (*G1* @*Bool* *y*)               = *ord y*
*f* (*G2* @*p* @*q x y z*)           = 0
*f* (*G3* @*p* @*q* @*a* @*Bool x y z*) = 1

In this definition
- *G1* has one type argument.
- *G2* has three type arguments, but we have chosen to match only the first two.
- *G3* is morally identical to *G2*, because of the equality, but it is written with four type arguments, and visible type application in patterns follows that specification.

### 4.3 Type aliases

In Section 3.5 we have seen that GHC currently restricts type variables to refer to type variables, but that this does not have to be the case. Similar questions arise in our function *f* above. Could we write this for *G3*?

*f* (*G3* @*p* @*q* @(*p*, *q*) @*b x y z*) = 1

Instead of *a* we have written (*p*, *q*), which is equal to *a*. And instead of *Bool* we have written *b*, thereby binding *b* to *Bool*.

$$\boxed{\Gamma \vdash e : \tau} \quad \text{Expression typing}$$

$$\frac{\Gamma, x : \tau_1 \vdash e : \tau_2}{\Gamma \vdash \lambda x.e : \tau_1 \to \tau_2} \; \text{Abs} \qquad \frac{\Gamma \vdash e_1 : \tau_1 \to \tau_2 \qquad \Gamma \vdash e_2 : \tau_2}{\Gamma \vdash e_1 \, e_2 : \tau_2} \; \text{App} \qquad \frac{\Gamma \vdash v : v \qquad \overline{c} = ftv(v)}{\Gamma, x : \forall \overline{c}. v \vdash e_2 : \tau}{\Gamma \vdash \mathbf{let} \; x :: v = e \; \mathbf{in} \; e_2 : \tau} \; \text{Let}$$

$$\frac{(v : \forall \overline{tv}. v) \in \Gamma}{\Gamma \vdash v : [\overline{tv \mapsto \tau}] v} \; \text{VarCon} \qquad \frac{\Gamma \vdash e : v \qquad \overline{\Gamma \vdash_p p_i : v \Rightarrow \Gamma_i' \quad \Gamma_i' \vdash e_i : \tau}^{\,i}}{\Gamma \vdash \mathbf{case} \; e \; \mathbf{of} \; \{\overline{p_i \to e_i}^{\,i}\} : \tau} \; \text{Case}$$

$$\boxed{\Gamma \vdash_p p : \sigma \Rightarrow \Gamma'} \quad \text{Pattern typing}$$

$$\frac{}{\Gamma \vdash_p x : \sigma \Rightarrow \Gamma, x : \sigma} \; \text{PatVar} \qquad \frac{(K : \forall \overline{a_i}^{\,i}. \overline{\sigma_k}^{\,k} \to T \, \overline{a_i}^{\,i}) \in \Gamma \qquad \overline{\Gamma \vdash_p^* \; \overline{p_k} : [\,\overline{a_i \mapsto \tau_i}^{\,i}\,] \sigma_k}^{\,k} \Rightarrow \Gamma'}{\Gamma \vdash_p K \, \overline{p_k}^{\,k} : T \, \overline{\tau_i}^{\,i} \Rightarrow \Gamma'} \; \text{PatCon98}$$

$$\boxed{\Gamma \vdash_p^* \; \overline{p_i : \sigma_i}^{\,i} \Rightarrow \Gamma'} \quad \text{Pattern sequence typing}$$

$$\frac{\overline{\Gamma_{i-1} \vdash_p p_i : \sigma_i \Rightarrow \Gamma_i}^{\,i \in 1..n}}{\Gamma_0 \vdash_p^* \; \overline{p_i : \sigma_i}^{\,i \in 1..n} \Rightarrow \Gamma_n} \; \text{PatSeq}$$

**Figure 3.** Typing of Haskell98 patterns

Given the ubiquity of equalities, it no longer seems to make sense to restrict what a scoped type variable can stand for, so we propose simply to drop the restriction. Doing so simplifies the specification and the implementation of both pattern signatures and visible type application in patterns. A GHC proposal by one of the authors [Breitner 2018] is underway. Relaxing the requirement also allows the user to use type variables as "local type synonyms" that stand for possibly long types:

*processMap* :: *Map Int* (*Maybe* (*Complex Type*)) → ...
*processMap* (*m* :: *Map key value*) = ...

## 5 Formal Specification

We give the first formal specification of a number of extensions to Haskell related to pattern matching and the scoping of type variables: annotations in patterns, scoped type variables, and type application syntax in patterns. This section builds up these specification step by step, starting with a specification of the language without these features.

Our specification does not cover **let**-bindings and declaration type signatures. We focus instead on pattern signatures, which is where our new contribution lies. The scoping of forall-bound type variables from declaration type signatures would be straightforward to add.

### 5.1 The baseline

We begin with a reduced model of Haskell98 terms, which knows nothing yet about scoped type variables nor type equalities (i.e., no GADTs). We also removed type class constraints. The syntax is given in Fig. 2, and the typing rules are in Fig. 3. In the typing rules, we use a convention where an over-bar indicates a list, optionally with a superscript index to indicate the iterator. Iterators are additionally annotated with length bounds, where appropriate.

The grammar includes separate metavariables for internal type variables $a$ and user type variables $c$. The former are type variables as propagated by the compiler, while the latter are type variables the user has written. It is as if internal type variables $a$ are spelled with characters unavailable in source Haskell. This distinction becomes important in Section 5.5. The language also includes only annotated **let**-bindings; no **let**-generalization here. (The "generalization" you might spot in Rule Let is simply quantifying over the variables the user has lexically written in the type signature.) This keeps our treatment simple and avoids the challenges of type inference. Allowing full **let**-generalization and un-annotated **let**s changes none of the conclusions presented here.

The judgment $\Gamma \vdash e : \tau$ indicates that the term $e$ has type $\tau$ in the context $\Gamma$, where $\Gamma$ is a list of term variables and their (possibly polymorphic) types. Data constructors are globally fixed in an initial context $\Gamma_0$; it is assumed that any context $\Gamma$ contains the global $\Gamma_0$ binding data constructors.

The type-checking of possibly nested patterns, as they occur in a case statement, is offloaded to the judgment $\Gamma \vdash_p p : \sigma \Rightarrow \Gamma'$, which checks that $p$ is a pattern for a value of type $\sigma$ and possibly binds new term variables, which are added to $\Gamma$ and returned in the extended environment $\Gamma'$. The auxiliary judgment $\Gamma \vdash_p^* \; \overline{p_i : \sigma_i}^{\,i} \Rightarrow \Gamma'$ straightforwardly threads the environment through a list of such pattern typings.

These rules should be unsurprising, but provide a baseline from which to build.

Updates to grammar:

| | |
|---|---|
| Constraints | $Q ::= \epsilon \mid Q_1 \wedge Q_2 \mid \tau_1 \sim \tau_2 \mid \ldots$ |
| Polytypes | $\sigma ::= \forall \overline{a}. Q \Rightarrow \tau$ |
| Type env | $\Gamma ::= \epsilon \mid \Gamma, v : \sigma \mid \Gamma, a : * \mid \Gamma, Q$ |
| Types of data cons | $\Gamma_0 = \overline{K : \forall \overline{a}. Q \Rightarrow \overline{\sigma} \rightarrow T \, \overline{\tau}}$ |
| Constraint entailment | $\Gamma \Vdash Q$ |

$\boxed{\Gamma \vdash e : \tau}$   Expression typing

$$\frac{(v : \forall \overline{a}. Q \Rightarrow v) \in \Gamma \qquad \Gamma \Vdash \overline{[\overline{a} \mapsto \tau]} Q}{\Gamma \vdash v : \overline{[\overline{a} \mapsto \tau]} v} \text{ VarConQ}$$

$$\frac{\Gamma \vdash e : \tau_1 \qquad \Gamma \Vdash \tau_1 \sim \tau}{\Gamma \vdash e : \tau} \text{ Eq}$$

$$\frac{\Gamma \vdash_p p_i : v \Rightarrow \Gamma'_i \qquad \Gamma'_i \vdash e_i : \tau^{\,i} \qquad \Gamma \vdash e : v \qquad ftv(\tau) \subseteq dom(\Gamma)}{\Gamma \vdash \textbf{case } e \textbf{ of } \{\overline{p_i \mapsto e_i}^{\,i}\} : \tau} \text{ CaseTv}$$

$$\frac{\Gamma, \overline{a : *} \vdash e : \overline{[\overline{c} \mapsto a]} v \qquad \overline{c} = ftv(v) \qquad \overline{c} \# dom(\Gamma) \qquad \Gamma, x : \forall \overline{c}. v \vdash e_2 : \tau}{\Gamma \vdash \textbf{let } x :: v = e \textbf{ in } e_2 : \tau} \text{ LetTv}$$

$\boxed{\Gamma \vdash_p p : \sigma \Rightarrow \Gamma'}$   Pattern typing

$$\frac{(\text{K} : \forall \overline{a}. Q \Rightarrow \overline{\sigma_i}^{\,i} \rightarrow T \, \overline{v_j}^{\,j}) \in \Gamma \qquad \overline{a} \# dom(\Gamma) \qquad \Gamma, \overline{a : *}, \overline{v_j \sim \tau_j}^{\,j}, Q \vdash_p \overline{p_i : \sigma_i}^{\,i} \Rightarrow \Gamma'}{\Gamma \vdash_p \text{K} \, \overline{p_i}^{\,i} : T \, \overline{\tau_j}^{\,j} \Rightarrow \Gamma'} \text{ PatCon}$$

**Figure 4.** Adding support for GADTs

## 5.2 Support for GADTs

Now we extend this language with support for GADTs, with their existential type variables and equality constraints. See Fig. 4. The term syntax is unchanged, but polytypes now can mention constraints, which can either be empty (and elided from this text), an equality between two monotypes, or a conjunction of constraints. We leave the possibility open for additional constraints, as indicated by the ellipsis.

The environment $\Gamma$ is extended with two new forms. First, we track the scope of type variables by adding $a : *$ to $\Gamma$[7]. Second, we add constraints $Q$ to $\Gamma$, to indicate which constraints (bound by a GADT pattern match) are in scope. Conversely, constraints are proved by the $\Gamma \Vdash Q$ entailment relation. As type inference and entailment is not the subject of this paper, we leave this relation abstract. The concrete instantiation of this judgment by, e.g., that of Vytiniotis et al. [2011] would be appropriate in an implementation of this type system.

Support for GADTs can be seen in the new form of data constructor types, listed in Fig. 4. Note that the arguments to

---

[7]Haskell supports higher kinds, but we elide that here for simplicity, and assume that all type variables have kind $*$.

T in the return type are no longer confined to be $\overline{a}$, the quantified type variables; instead they can be arbitrary monotypes. In addition, a constructor can include a constraint $Q$.

When a data constructor is used in an *expression*, then the type equalities must be satisfied in the current environment, as expressed by the new premise of Rule VarConQ in Fig. 4. We see also that the type equalities in the environment can be used for implicit coercion, as expressed in the Rule Eq.

When *pattern-matching* a data constructor, Rule PatCon brings the type variables $\overline{a}$ into scope, by extending $\Gamma$. We require that these bound variables are fresh with respect to other variables in scope, a requirement we can satisfy by $\alpha$-renaming if necessary. We also add the type equalities that we have learned to the environment—that is, the equivalence between the $\overline{v_j}^{\,j}$ from the data constructor's type and the $\overline{\tau_j}^{\,j}$ from the pattern type.

Finally, we update Rule CaseTv to prevent skolem escape and Rule LetTv to track the internal variables brought into scope. Note that these variables are *internal* only—the user cannot write them in a program.

At this point, our type system is comparable in expressiveness to the specification given by Vytiniotis et al. [2011]. A notable difference is that we explicitly handle nested patterns. This is important, as in the presence of GADTs, the precise formulation of how nested pattern are type-checked matters. For example, consider:

```
data G a where
    G1 :: G Bool
    G2 :: G a

f :: (G a, a, G a) → Bool
f (G1, True, _) = False
f (_, True, G1) = False
```

Here the first equation for $f$ is fine, but the second is not, because the pattern *True* cannot match against an argument of type $a$ until *after* the constructor *G1* has been matched—and matching in Haskell is left-to-right.

## 5.3 Treating universals and existentals uniformly

A technical contribution of this paper is that Rule PatCon is simpler and more uniform than the one usually given [e.g. by Vytiniotis et al. 2011], in that *it does not distinguish the universal and existential type variables of the data constructor*. Instead, *all* the type variables are freshly bound, with the equalities $\overline{v_j \sim \tau_j}^{\,j}$ linking them to the context. In particular, these equalities take the place of the substitution written in the previous Rule Con98.

However, there is a worry: pattern-matching involving GADTs lacks principal types, and hence usually requires a type signature (see Section 2.1). If we treat vanilla, non-GADT Haskell98 data types in the same way as GADTs, do we lose type inference for ordinary Haskell98 definitions? Specifically, Vytiniotis et al. [2011, Section 5.6.1] describe

Patterns                                    $p ::= \ldots \mid p :: \sigma$

$$\frac{ftv(\sigma') = \emptyset \qquad \Gamma \Vdash \sigma \leq \sigma' \qquad \Gamma \vdash_{\mathrm{p}} p : \sigma' \Rightarrow \Gamma'}{\Gamma \vdash_{\mathrm{p}} (p :: \sigma') : \sigma \Rightarrow \Gamma'} \text{ PatSig}$$

**Figure 5.** Syntax and typing rule for pattern signatures

how assumed local constraints can interfere with type inference, essentially by making certain unification variables "untouchable" (that is, unavailable for unification). That section also describes how to make more unification variables touchable in the non-GADT case, when the constraints entail no equalities. But *our typing rule introduces equalities even in the non-GADT case*, so this mechanism fails for us.

Let us investigate Rule PatCon specialized to the case of an ordinary, non-GADT constructor, which binds no context $Q$ and does not constrain its result type arguments:

$$\frac{(K : \forall \overline{a_j}^{\,j}.\, \overline{\sigma_i}^{\,i} \rightarrow T\, \overline{a_j}^{\,j}) \in \Gamma \qquad \overline{a} \,\#\, dom(\Gamma)}{\Gamma, \overline{a_j : *}^{\,j},\, \overline{a_j \sim \tau_j}^{\,j} \vdash_{\mathrm{p}}^{*} \overline{p_i : \sigma_i}^{\,i} \Rightarrow \Gamma'}{\Gamma \vdash_{\mathrm{p}} K\, \overline{p_i}^{\,i} : T\, \overline{\tau_j}^{\,j} \Rightarrow \Gamma'} \text{ PatCon98'}$$

We see that all of its assumed equality constraints take the form $a_j \sim \tau_j$, where $a_j$ is freshly bound. We can view such equalities not as true assumed equalities (which lead to the type inference problems for GADTs), but instead as a form of local **let**-binding: the context simply gives us the definition of these type variables. In this interpretation, it is critical that the type variable in the equality assumption is freshly bound—that is, we are not referring to a type variable from a larger scope. Viewing the equalities in $\Gamma'$ as **let**-like, it is sensible to extend the ad-hoc extension of Vytiniotis et al. [2011] to include such forms. Indeed, doing so is an independently-useful improvement to type inference, and GHC has already adopted it, in response to a request[8] from one of this paper's authors. Thus, despite the addition of equalities in Rule PatCon, we do not have a negative effect on type inference.

### 5.4 Closed pattern signatures

Our next step is to formalize *PatternSignatures*, which allows the user to annotate patterns with type signatures, but for now we will only handle *closed* pattern signatures. We simply add one new typing rule PatSig, shown in Fig. 5. Note that the user is allowed to give the pattern a more specific type, as in this example, which requires *RankNTypes*:

*f* :: $(\forall a.\, a \rightarrow a) \rightarrow Int$
*f* $(x :: Int \rightarrow Int) = x\ 42$

The typing rule expresses this through the premise $\Gamma \Vdash \sigma \leq \sigma'$, an appeal to the subtype relation on polytypes. This subtype relationship checks that the expected type of the pattern $\sigma$ is more general than the annotated type $\sigma'$. Note that this

$$\frac{\sigma = \forall \overline{c}.\, Q \Rightarrow \upsilon \qquad ftv(\sigma) \subseteq dom(\Gamma)}{\Gamma, \overline{c : *},\, Q \vdash e : \upsilon \qquad \Gamma, x : \sigma \vdash e_2 : \tau}{\Gamma \vdash \mathbf{let}\ x :: \sigma = e\ \mathbf{in}\ e_2 : \tau} \text{ LetForall}$$

$$\frac{\overline{c} = ftv(\sigma') \setminus dom(\Gamma) \qquad \Gamma' = \Gamma, \overline{c : *},\, \overline{c \sim b}}{\Gamma' \Vdash \sigma \leq \sigma' \qquad \Gamma' \vdash_{\mathrm{p}} p : \sigma' \Rightarrow \Gamma''}{\Gamma \vdash_{\mathrm{p}} (p :: \sigma') : \sigma \Rightarrow \Gamma''} \text{ PatSigTv}$$

**Figure 6.** Typing with scoped type variables

relationship is *backwards* from the usual expected/actual relationship in typing because patterns are in a negative position. The subtleties of of polytype subtyping are well explored in the literature[9] and need not derail our exploration here. However, note that Rule PatSig checks pattern $p$ with the annotated type $\sigma'$, not the more general $\sigma$—after all, the user has asked us to use $\sigma'$.

### 5.5 Scoped type variables

Next, we add support for *open* pattern signatures that bring type variables into scope. We need not change any syntax. Only the typing rule for annotated patterns changes: we replace PatSig with PatSigTv and add LetForall in Fig. 6.

The LetForall rule allows programmers to bring variables $\overline{c}$ into scope when an explicit $\forall$ is mentioned in the source. Note that this rule does not do any implicit lexical generalization: echoing GHC's behavior, if the user writes a $\forall$, *all* new variables to be used in the type signature must be bound explicitly.

In Rule PatSigTv, the last two premises are identical to those of its predecessor PatSig. The first premise extracts the type variables $\overline{c}$ that are free in the user-written type signature $\sigma'$, but not already in scope in $\Gamma$. The "not already in scope" part reflects the discussion of Section 3.1.

But what if $\Gamma$ contains a binding, introduced by rule PatCon, for a type variable that just happens to have the same name as one of the $\overline{c}$ in a user-written signature? After all, the names of the type variables in PatCon are arbitrary internal names; they just need to be fresh. Our solution is simple: we take advantage of the difference between internal type variables and external ones. The user cannot accidentally capture an internal variable.

The second (top-right) premise of Rule PatSigTv is the most unusual. It brings the variables $\overline{c}$ into scope, but then *also* assumes that each variable $c$ equals some internal variable $b$. (Nothing stops multiple elements of $\overline{b}$ from being the same internal variable.) Strikingly, the $\overline{b}$ are mentioned nowhere else in the rule. The existence of the $\overline{b}$ in this premise essentially says that the $\overline{c}$ are merely a renaming of existing in-scope internal variables. In practice, the $\overline{b}$ are chosen in order to make the subtyping relationship $\Gamma' \Vdash \sigma \leq \sigma'$

Patterns $\qquad p ::= x \mid K\ \overline{@\tau}\ \overline{p} \mid p :: \sigma$

$$\frac{\begin{array}{c}(K : \forall \overline{a_j}^{\,j}.\, Q \Rightarrow \overline{\sigma_k}^{\,k} \to T\ \overline{v_i}^{\,i})\ \in\ \Gamma \\ \overline{c_l}^{\,l} = \mathit{ftv}(\overline{\tau_j'}^{\,j}) \setminus \mathit{dom}(\Gamma) \qquad \overline{a_j}^{\,j}\ \#\ \mathit{dom}(\Gamma) \\ \Gamma' = \Gamma, \overline{a_j : *}^{\,j}, \overline{c_l : *}^{\,l}, \overline{c_l \sim b_l}^{\,l}, \overline{v_i \sim \tau_i}^{\,i}, Q \\ \overline{\Gamma' \Vdash \tau_j' \sim a_j}^{\,j} \\ \Gamma' \vdash_{\mathrm{p}}^{+} \overline{p_k : \sigma_k}^{\,k} \Rightarrow \Gamma'' \end{array}}{\Gamma \vdash_{\mathrm{p}} K\ \overline{@\tau_j'}^{\,j}\ \overline{p_k}^{\,k} : T\ \overline{\tau_i}^{\,i} \Rightarrow \Gamma''}\ \textsc{PatConTyApp}$$

**Figure 7.** Typing of type applications in patterns

hold; GHC checks this subtyping relationship, unifying the $\overline{c}$ with internal variables $\overline{b}$ as necessary. Because the subtyping relationship is checked with respect to a context that contains the $\overline{c \sim b}$ equalities, the $\overline{b}$ do not need to be explicitly mentioned again in the rule. For example, consider

```
data ExIntegral where    -- packs an Integral value
   MkEx :: ∀a. Integral a ⇒ a → ExIntegral

getInt :: ExIntegral → Integer
getInt (MkEx (x :: c)) = toInteger @c x
```

The pattern match on *MkEx* brings an internal existential variable *a* into scope, via the PatCon rule. Recall that the user cannot type the name of such a variable. Instead, the user annotates the pattern *x* with the user-written type variable *c*. This annotation triggers Rule PatSigTv, which must find an internal variable *b* such that $a : *, c : *, c \sim b \Vdash a \leq c$. The answer is that we must choose *b* to be equal to the variable *a*, and the rule succeeds. We have thus renamed the internal variable *a* to become the user-written variable *c* and can successfully use *c* in the pattern's right-hand side.

Contrast that behavior with this (failing) example:

```
notAVar :: Int → Int
notAVar (x :: c) = x
```

Here, we are trying to bind a user-written type variable *c* to *Int*. GHC rejects this function, saying that *c* does not match with *Int*. In terms of Rule SigPatTv, there exists no *b* such that $c : *, c \sim b \Vdash Int \leq c$ holds.

There is a free design choice embodied in Rule SigPatTv: the rule asserts that each *c* must be a renaming of a type *variable*. Instead, we could replace $\overline{c \sim b}$ with $\overline{c \sim \tau}$, allowing each type variable to rename a *type*. Nothing else in the system would have to change. Indeed, understanding this very fact is one of the primary motivators for writing this specification in the first place.

### 5.6 Type applications in patterns

Having nailed down the status quo, it is now easy to specify what it should mean to use type applications in patterns.

This version supports type applications only in *constructor* patterns; we study pattern synonyms in Appendix A. The syntax and new typing rule are shown in Fig. 7. Rule PatConTyApp looks scary, but it just integrates the concepts seen in Rule PatSigTv into Rule PatCon. We have kept all the iteration indices to help the reader match up which lists are expected to have the same size.

Let us look at each premise separately:

- Once again, the type variables $\overline{c_l}^{\,l}$ are those that occur in the explicit type patterns but are not yet in scope. These are treated like type variables in a pattern signature: they are brought into scope here, each as a short-hand for some internal type variable $b_l$.
- The environment $\Gamma$ is extended to $\Gamma'$ and contains now the (internal) type variables $\overline{a_j}^{\,j}$, the user-written scoped type variables $\overline{c_l}^{\,l}$, the type equations equating each $c_l$ to its internal type variable $b_l$, the GADT equalities $\overline{v_i \sim \tau_i}^{\,i}$, and the constraint $Q$ captured by K.
- The type patterns are checked against the types they match against. In contrast to pattern signatures, we use type equality here ($\sim$), not the subtyping relation ($\leq$): no types involved can be polytypes, and so the subtyping relation degenerates to type equality.

As written here, the rule requires a type application for each type variable (note that the $\overline{@\tau_j'}^{\,j}$ use the same indices as the quantified type variables $\overline{a_j}^{\,j}$ in K's type). However, we can weaken this requirement simply by dropping some $\tau_j'$s from the conclusion and the relevant premises.

Just as in Rule SigPatTv:

- The $\overline{b_l}^{\,l}$ are mentioned nowhere else in the rule; instead, they are fixed such that the equality constraints for the $\overline{\tau_j'}^{\,j}$ are entailed by $\Gamma'$.
- The rule requires that each user-written type variable stands for an internal *variable*, but this choice could readily be changed by writing $\overline{v_l'}^{\,l}$ instead of $\overline{b_l}^{\,l}$ in the definition of $\Gamma'$.

### 5.7 Conclusion

Through the incremental building of rules, we can see precisely how the new feature of explicit binding sites for type variables fits into the existing typing framework. We have also explored two further extensions:

- Allowing type application in patterns headed by *pattern synonym* [Pickering et al. 2016]. Our framework extends well in this new context, offering no surprises (Appendix A).
- Incorporating explicit binding sites for type variables in the patterns of a $\lambda$-expression. This is slightly subtler (though the end result adds only one, simple typing rule), but is relegated to the appendix because it requires reasoning about bidirectional type checking.

Bringing all the necessary context into scope would take us too far afield here (Appendix B).

## 6 Alternative approaches

### 6.1 Universals vs. existentials

Type theorists are wont to separate quantified type variables in data constructors into two camps: *universals* and *existentials*. Here is a contrived but simple example:

```
data UnivEx a where
  MkUE :: ∀a b. a → b → UnivEx a

matchUE :: ∀a. UnivEx a → ...
matchUE (MkUE x y) = ...
```

In the constructor *MkUE*, the variable *a* is universal (it is fixed by the return type *UnivEx a*) while *b* is existential (it is not fixed by the result type). When we match on *MkUE* in *matchUE*, we might want to bind *b*, as it is first brought into scope by the match. However, we never need to match *a*, as it is already in scope from *matchUE*'s type signature.

An alternative design for type applications in patterns is to allow matching only existentials in pattern matches, thus:

```
matchUE :: ∀a. UnivEx a → ....
matchUE (MkUE @b x y) = ...
```

Indeed, this forms the main payload of the original GHC proposal for binding type variables [Suarez 2017]. This design is attractive because the bindings would be concise: only those variables that need to be bound would be available. However, there are two distinct drawbacks:

- Given the complexity of Haskell, it can be hard to specify the difference between universals and existentials: Clearly, *a* is universal in the constructor *MkUE* above. But what if its type were *MkUE* :: ∀ *a b. a → b → UnivEx* (*Id a*), where *Id* is a type synonym? An injective type family? If we add *a ~ b* to the constraints of *MkUE*, then *b* is also fixed by the result type—does that make it a universal?
  The question of whether the value of a type variable is fixed by the return type depends on how smart the compiler is, and any specification would have to draw an arbitrary line. In the end, this would leave our users just very confused.
- When using a data constructor in an *expression*, the caller is free to instantiate both universals and existentials. Indeed, universals and existentials are utterly indistinguishable in expressions. That means that one might write *MkUE* @*Int* @*Bool* 5 *True* in an expression. If we could match against only existentials in patterns, though, we would have to write a pattern *MkUE* @*b x y*, remembering to skip the universal *a*. This would both be confusing to users and weaken the ergonomics of pattern matching, whose chief virtue is that deconstructing a datatype resembles closely the syntax of constructing one.

We thus prefer not to differentiate universals and existentials in this way.

### 6.2 The type-lambda approach

A plausible alternative approach to adding scoped type variables is to take a hint from System F, the explicitly-typed polymorphic lambda calculus [Girard 1990]. In System F, a *type lambda*, written "Λ", binds a type variable, just as a term lambda, written "λ", binds a term variable. For example:

$$id : \forall \alpha. \ \alpha \to \alpha$$
$$id = \Lambda \alpha. \ \lambda x : \alpha. \ x$$

A term $\Lambda \alpha.e$ has type $\forall \alpha.\tau$, for some type $\tau$, just as a term $\lambda x.e$ has type $\tau_1 \to \tau_2$. Hence, a very natural idea is to bind a source-language type variable with a source-language type lambda. This "the type-lambda approach" is the one adopted by SML 97 [Milner et al. 1997]. In SML one can write:

```
fun 'a prefix (x : 'a) yss =
  let fun xcons (ys : 'a list) = x :: ys in
  map xcons yss
```

Here, "'a" following the keyword fun is the binding site of an (optional) type parameter of prefix; it scopes over the patterns of the definition and its right hand side.

Just as Haskell has implicit quantification in type signatures, SML allows the programmer to introduce *implicit* type lambdas. This definition is elaborated into the previous one:

```
fun prefix (x : 'a) yss =
  let fun xcons (ys : 'a list) = x :: ys in
  map xcons yss
```

The language definition gives somewhat intricate rules to explain how to place the implicit lambdas. For example:

```
fun f x = ....(fun (y:'a) => y)....
```

Where is the type lambda that binds the type variable 'a? In SML one cannot answer that question without knowing both what the "...." is, and the context for the definition fun f. Roughly speaking, the type lambda for an implicitly-scoped type variable 'a is placed on the innermost function definition that encloses all the free occurrences of 'a. The rule [Milner et al. 1997] is only one informal, albeit carefully worded, paragraph; the formal typing rules assume that a pre-processing pass has inserted an explicit binding for every type variable that is implicitly bound by the above rule.

The type-lambda approach explicitly connects *lexical scoping* and *quantification*. In contrast, our approach decouples the two, by treating a lexically scoped type variable merely as an alias for a type (or type variable).

## Acknowledgments

This material is based upon work supported by the National Science Foundation under Grant No. 1319880, Grant No. 1521539, and Grant No. 1704041.

| Pattern synonyms | | $\ni P$ |
|---|---|---|
| Expressions | $e ::=$ | $\ldots \mid P$ |
| Patterns | $p ::=$ | $\ldots \mid P \, \overline{@\tau} \, \overline{p}$ |
| Pattern synonym types | $\Gamma_1 = P : \forall \overline{a}. \, Q_r \Rightarrow \forall \overline{b}. \, Q_p \Rightarrow \overline{\sigma} \rightarrow v$ | |

$$\boxed{\Gamma \vdash e : \tau} \quad \text{Expression typing}$$
$$(P : \forall \overline{a}. \, Q_r \Rightarrow \forall \overline{b}. \, Q_p \Rightarrow v) \in \Gamma$$
$$\theta = [\overline{a \mapsto \tau}] \, [\overline{b \mapsto \tau'}]$$
$$\dfrac{\Gamma \Vdash \theta(Q_r \wedge Q_p)}{\Gamma \vdash P : \theta(v)} \quad \textsc{Syn}$$

$$\boxed{\Gamma \vdash_p p : \sigma \Rightarrow \Gamma'} \quad \text{Pattern typing}$$
$$(P : \forall \overline{a_i}^{\,i}. \, Q_r \Rightarrow \forall \overline{b_j}^{\,j}. \, Q_p \Rightarrow \overline{\sigma_k}^{\,k} \rightarrow v) \in \Gamma$$
$$\Gamma' = \Gamma, \overline{a_i : *}^{\,i}, \overline{a_i \sim v_i'}^{\,i} \qquad \overline{a_i}^{\,i} \, \# \, dom(\Gamma)$$
$$\Gamma' \Vdash (v \sim \tau) \wedge Q_r$$
$$\overline{c_l}^{\,l} = ftv(\overline{\tau_i'}^{\,i}, \, \overline{\tau_j''}^{\,j}) \setminus dom(\Gamma) \qquad \overline{b_j}^{\,j} \, \# \, dom(\Gamma)$$
$$\Gamma'' = \Gamma', \overline{b_j : *}^{\,j}, \overline{c_l : *}^{\,l}, c_l \sim \overline{b_l'}^{\,l}, Q_p$$
$$\overline{\Gamma'' \Vdash \tau_i' \sim a_i}^{\,i}$$
$$\overline{\Gamma'' \Vdash \tau_j'' \sim b_j}^{\,j}$$
$$\dfrac{\overline{\Gamma'' \vdash_p p_k : \theta(\sigma_k)}^{\,k} \Rightarrow \Gamma'''}{\Gamma \vdash_p P \, \overline{@\tau_i'}^{\,i} \, \overline{@\tau_j''}^{\,j} \, \overline{p_k}^{\,k} : \tau \Rightarrow \Gamma'''} \quad \textsc{PatSyn}$$

**Figure 8.** Support for pattern synonyms

## A  Pattern synonyms

Pattern synonym [Pickering et al. 2016] types differentiate universal type variables from existential type variables, and thus add superficial complexity to our rules. Happily, this complexity really is just superficial—the underlying mechanism we have laid out applies well in this new scenario. See Fig. 8.

A pattern synonym type has *two* contexts: the $Q_r$ *required* constraint and the $Q_p$ *provided* constraint. Using a pattern synonym in a pattern requires that the required constraint already be provable; in contrast, the provided constraint may be assumed in the pattern's right-hand side. The provided constraint is the analogue of the constraint in a data constructor's type, while the required constraint has no data constructor analogue. An example can help to demonstrate. If we define **pattern** *Three* = 3 and then write

*f* :: *T* → ...
*f Three* = ...

we must be able to prove both that *Num T* and *Eq T* hold before the pattern match can type-check. Accordingly, the pattern *Three* gets type $\forall a. \, (Eq \; a, \, Num \; a) \Rightarrow a$, where the constraint listed is the required one. (When only one constraint is written, we understand that it is the required one, not the provided one.)

Because of the two contexts, our old Rule VarConQ is insufficient to deal with pattern synonyms used in an expression. We thus introduce the straightforward Rule Syn.

Corresponding to the two contexts, a pattern synonym has two sets of type variables, the universals $\overline{a_i}^{\,i}$ and the existentials $\overline{b_j}^{\,j}$. The universals are those that are mentioned in the result type $v$; they can be fixed by comparing $v$ with $\tau$, the known type of the pattern[10]. In contrast, the existentials are simply brought into scope by the pattern; they match no other known type (except as entailed by the constraints $Q_p$).

Let us work premise-by-premise through Rule PatSyn.

- We first look up the pattern synonym type in the environment.
- We then bring the universals $\overline{a_i}^{\,i}$ into scope, allowing each universal variable $a_i$ to equal some other type $v_i'$. We can think of these $v_i'$ as the instantiations for the variables $a_i$.
- Under these assumptions (that is, after instantiating universals with arbitrary monotypes $\overline{v_i'}^{\,i}$), we must ensure that the expected type of the pattern $\tau$ matches the declared result type of the pattern synonym $v$. We also must ensure that the required constraint $Q_r$ indeed holds.
- Then, we collect the newly scoped type variables $\overline{c_l}^{\,l}$.
- We extend the context yet again, bringing the existentials $\overline{b_j}^{\,j}$ into scope, introducing the user-written scoped type variables $\overline{c_l}^{\,l}$, asserting that each $c_l$ is actually a renaming of an internal variable $b_l'$, and assuming the provided constraint $Q_p$. (As before, we could reasonably change the $b_l'$ to a $v_l''$ to allow scoped type variables to stand for types instead of variables.)
- We now check that the type patterns corresponding to universals $\overline{\tau_i'}^{\,i}$ indeed match the universals $\overline{a_i}^{\,i}$.
- We also check that the type patterns corresponding to existentials $\overline{\tau_j''}^{\,j}$ match those existentials $\overline{b_j}^{\,j}$.
- Finally, we use this extended context to check the pattern arguments $\overline{p_k}^{\,k}$.

Though there are lots of details here, the only new piece is the check for the required constraint $Q_r$. This new check fits nicely within our framework.

---

[10] There are some subtleties around the difference between universals and existentials. Pickering et al. [2016, Section 6.3] explore these challenges.

# B  Binding type variables in function definitions

## B.1  Motivation and examples

It is relatively easy to extend the ideas in our paper to allow type variable binding in function definitions. We can motivate this idea using an early example, repeated here:

*prefix* :: $\forall a.\ a \rightarrow [[a]] \rightarrow [[a]]$
*prefix x yss = map xcons yss*
   **where** *xcons* :: $[a] \rightarrow [a]$
       *xcons ys = x : ys*

The type signature includes a $\forall$ which brings the type variable $a$ into scope in the body of *prefix*. We find this syntax design awkward however: a traditional reading of the syntax of $\forall$ tells us that $a$ is in scope in $a \rightarrow [[a]] \rightarrow [[a]]$ and nowhere else. Indeed, we should be able to $\alpha$-vary the name $a$ in that type signature (and only that type signature) without any external effect. Yet, it is not so, with $a$'s scope extending beyond that type signature.

Instead, we find it would be more natural (if, admittedly, a bit more verbose) to write

*prefix* @$a$ *x yss* = ...

explicitly binding the type variable $a$ in a pattern.

**GeneralizedNewtypeDeriving**  This issue is not always so simple, however. Consider GHC's implementation of the extension *GeneralizedNewtypeDeriving*, which uses *coerce* [Breitner et al. 2016] to translate the implementation of a class method from one type to another. Here is an example:

**newtype** *MyList a = MkMyList* $[a]$
   **deriving** *Functor*

How should we write our derived *Functor* instance? A first attempt might look like this:

**instance** *Functor MyList* **where**
   *fmap = coerce* (*fmap* @$[\,]$)

This, unfortunately, fails to compile. The problem is that the use of *coerce* is ambiguous. GHC can determine that the argument to *coerce* has type $(a0 \rightarrow b0) \rightarrow [a0] \rightarrow [b0]$ (the lists in there come from the use of @$[\,]$) and that the result must have type $(a \rightarrow b) \rightarrow MyList\ a \rightarrow MyList\ b$. However, GHC has no way to know that the undetermined $a0$ and $b0$ in *coerce*'s argument's type should be the expected $a$ and $b$ in the result type. Perhaps GHC could guess this correspondence, but GHC refuses to make guesses.

The way to get this to work is to bring $a$ and $b$ into scope, like this:

**instance** *Functor MyList* **where**
   *fmap* :: $\forall a\ b.\ (a \rightarrow b) \rightarrow MyList\ a \rightarrow MyList\ b$
   *fmap = coerce* (*fmap* @$[\,]$ @$a$ @$b$)

Expressions                          $e ::= \ldots \mid \lambda @c.e$

$$\boxed{\Gamma \vdash^*_{sb} e \Leftarrow \sigma}$$  Checking against specified polytypes

$$\frac{\Gamma, c : * \vdash^*_{sb} e \Leftarrow [a \mapsto c]\sigma}{\Gamma \vdash^*_{sb} \lambda @c.e \Leftarrow \forall a.\ \sigma}\ \text{SB\_DTyAbs}$$

**Figure 9.** Binding type variables in $\lambda$-expressions

By writing a type signature for *fmap* (using the *InstanceSigs* extension), we can bring $a$ and $b$ into scope to use as visible type arguments to the coerced *fmap*. As with the examples in our motivation section, this workaround is infelicitous, requiring enabling yet another extension and writing out an otherwise-redundant type, just to bind type variables.

With the ability to explicitly bind type variables, this instance is simplified to become

**instance** *Functor MyList* **where**
   *fmap* @$a$ @$b$ = *coerce* (*fmap* @$[\,]$ @$a$ @$b$)

**RankNTypes**  Further motivation arises in the context of higher-rank types [Peyton Jones et al. 2007]. Suppose we have

*hr* :: $(\forall a.\ F\ a \rightarrow F\ a) \rightarrow \ldots$

where $F$ is a non-injective type family. At a call site of *hr*, ...*hr* $(\lambda x \rightarrow \ldots)$..., there is no way to bind the type variable $a$, which is used only as the argument to non-injective type family. Currently, the only workaround here is to add a *Proxy* argument to *hr*'s argument type, allowing the client of *hr* to bind the type variable in a pattern annotation on the proxy argument.

Instead, our design allows users to bind the type variable explicitly, with ...*hr* $(\lambda @a\ x \rightarrow \ldots)$....

## B.2  Specification

The ability to bind type variables makes sense only when we know the type of the function we are defining; this type tells us how the type variable relates to the other parameters and the result of the function. Note that the analogue of this requirement is always true for constructor patterns, where we can look up the type of the constructor in the environment. We thus must proceed by considering a bidirectional approach to type-checking, where explicitly track whether we are *inferring* a type or *checking* one. Specifically, we will build on one author's previous work [Eisenberg et al. 2016], which sets the stage for our new extension here. Readers may want to consult Fig. 8 of that previous work in order to understand our new contribution.

Our addition to the previous $\vdash^*_{sb}$ judgment is in Fig. 9. The $\vdash^*_{sb}$ judgment checks an expression against a polytype containing outermost binders. The original judgment comprises only one rule, which skolemizes all these variables into fresh type constants. With our addition, it is understood that the original rule, termed SB_DEEPSKOL, fires only when the new one does not apply. Our new rule simply brings $c$ into scope as an alias for the internal variable $a$ from the polytype we are checking against.

Incredibly, that's it! No other changes are necessary. One of the claims in the previous work is that the design therein is flexible and extensible. Our success here is further evidence of that claim.

There are a few subtleties of this approach, of course:

**Type variables vs. types**   The grammar extension and Rule SB_DTYABS both require that the $\lambda$-expression binds a *variable*, never a full type. This contrasts with our approach for binding type variables in constructor patterns. The reason for the difference here is that, while we could mimic the earlier approach here, we would find that the type pattern would *always* just be a variable. In the case of $\lambda$-expressions, there is no analogue of the universal variables of data constructor types. Furthermore, any context $Q$ in the polytype being checked against will come into scope *after* the bound type variable, so there is no possibility of such a context entailing an equality between the bound variable and some other type. For similar reasons, there is no incentive to use an equality assumption in place of substitution in Rule SB_DTYABS. We thus keep the rule simple by straightforwardly requiring the binding of only variables, and not full types.

**Comparison against declarative specification**   Since we are working in the context of Eisenberg et al. [2016], we should consider what the implications are for the comparison of our extension of System SB against the declarative System B of that work.[11] Happily, all we need to do to keep System B in sync with System SB is add an analogous rule to the "checking" judgment of System B: the fact that the judgment works only with *specified* polytypes (never *generalized* ones, in the terminology of the previous work) means that we need not worry about the effects of implicit generalization here.

**Deep skolemization**   Unfortunately, the new system is not quite as expressive as we would like. Rule SB_DEEPSKOL performs *deep* skolemization [Peyton Jones et al. 2007, Section 4.6], meaning that any quantified type variables—even those to the right of arrows—are skolemized. Once these are skolemized, they are unavailable in later applications of SB_DTYABS. Let us examine an example:

$(\lambda @c1 \ x \ y \rightarrow \ldots) :: \forall a1. \ a1 \rightarrow \forall a2. \ a2 \rightarrow \ldots$

In this example, we are checking a $\lambda$-expression binding type variables against a polytype $\forall a1. \ a1 \rightarrow \forall a2. \ a2 \rightarrow \ldots$. First, SB_DTYABS applies, binding $c1$ to be an alias for $a1$. Then, however, SB_DEEPSKOL applies, which will skolemizes $a2$, removing an opportunity to bind it. Accordingly, while the example above is accepted by our addition, this one would be rejected:

$(\lambda @c1 \ x \ @c2 \ y \rightarrow \ldots) :: \forall a1. \ a1 \rightarrow \forall a2. \ a2 \rightarrow \ldots$

The difference here is that we try to bind $c2$ as an alias to $a2$. This fails because $a2$ has been skolemized before we can apply SB_DTYABS.

One could easily wonder: why do deep skolemization here? This is answered in the last paragraph of Section 6.1 of the extended version of Eisenberg et al. [2016]—essentially, doing shallow skolemization would work but would make our ideas harder to implement. Currently, we have no non-contrived examples where this extra work is necessary, and so we leave the deep skolemization rule as it is.

---

[11]The previous work builds up its type systems by contrasting syntax-directed type systems against declarative specifications. The current paper focuses only on the syntax-directed approach.